\documentclass[10pt,journal]{IEEEtran}

\usepackage[center]{caption}

\usepackage{xcolor}

\usepackage{cite}

\ifCLASSINFOpdf
\usepackage[pdftex]{graphicx}
\graphicspath{{img/}}
\else
\fi

\usepackage{amsmath}

\usepackage{algorithmic}
\usepackage{algorithm}
\usepackage{makecell}
\usepackage{siunitx, mhchem}

\usepackage{stfloats}

\begin{document} 
\pagenumbering{gobble}	
\title{SensorDrop: A Reinforcement Learning Framework for Communication Overhead Reduction on the Edge}

\author{

\IEEEauthorblockN{Pooya Khandel\IEEEauthorrefmark{1},
Amir Hossein Rassafi\IEEEauthorrefmark{1},
Vahid Pourahmadi\IEEEauthorrefmark{1}, Saeed Sharifian\IEEEauthorrefmark{1},
and Rong Zheng\IEEEauthorrefmark{2}} 

\IEEEauthorblockA{\IEEEauthorrefmark{1}
Department of Electrical Engineering, Amirkabir University of Technology,
Tehran, Iran \\}
\IEEEauthorblockA{\IEEEauthorrefmark{2} Department of Computing and Software,
McMaster University, Hamilton, ON, Canada} 
}

	\maketitle
	
	\begin{abstract}
	In IoT solutions, it is usually desirable to collect data from  a large number of distributed IoT sensors at a central node in the cloud for further processing. One of the main design challenges of such solutions is the high communication overhead between the sensors and the central node (especially for multimedia data). In this paper, we aim to reduce the communication overhead and propose a method that is able to determine which sensors should send their data to the central node and which to drop data. The idea is that some sensors may have data which are correlated with others and some may have data that are not essential for the operation to be performed at the central node. As such decisions are application dependent and may change over time, they should be learned during the operation of the system, for that we propose a method based on Advantage Actor-Critic (A2C) reinforcement learning which gradually learns which sensor's data is cost-effective to be sent to the central node. The proposed approach has been evaluated on a multi-view multi-camera dataset, and we observe a significant reduction in communication overhead with marginal degradation in object classification accuracy.
	\end{abstract}
	
	
	\IEEEpeerreviewmaketitle

	\section{Introduction}

Internet of Things (IoT) has attracted much research interest in recent years.
It is predicted that there will be around 50 billion IoT devices 2020
\cite{gambirovza2018big}. In IoT applications, it is common to collect a large
amount of information from distributed sensors and perform a complex task, e.g.,
executing a Machine Learning (ML) inference model over the data collected. 
The key requirements for IoT applications include accuracy, latency and power
consumption on IoT solutions. 

Currently, there are two main paradigms for IoT solutions, namely, fully
decentralized and cloud-based.  In fully decentralized solutions, sensors are
equipped with processing units and can perform basic inference tasks. To reduce
the computation complexity, low complexity ML models have been investigated in
literature. 
Quantized CNN \cite{wu2016quantized} reduces computational
complexity by quantizing the filter weights of convolutional layers. Huang and
Wang \cite{huang2018data} propose a pruning mechanism to remove a few of
unnecessary connections and/or layers of a deep neural network. BlockDrop
\cite{wu2018blockdrop} focuses on residual networks (ResNet)
and  suggests to drop extra and unnecessary blocks of ResNet for a given input
image. The authors show that it is possible to achieve almost the same accuracy
as a complete ResNet model even though not all residual blocks are utilized.
In cloud-based solutions, sensor data are collected at a cloud node (CN) for
centralized processing. The cloud-based solutions are attractive as they are
not generally limited by computation resources and they can take advantage of
the global view of all sensors for the inference tasks. However, such solutions
suffer from high communication overhead in transferring data from distributed
sensors and potentially high latency.  

To mitigate these problems in
cloud-based solutions, building upon  their earlier work on BranchyNet
\cite{teerapittayanon2016branchynet}, Teerapittayanon {\it et
al.}~\cite{teerapittayanon2017distributed} propose a new distributed deep
neural networks architecture (DDNNs) for cloud-based ML. In DDNN, before
sending data to the CN, sensors first send their data to an edge device, called
{\it local aggregator} (LA), which checks to see if it can perform the task
itself. If LA is successful, it saves both time and communication compared to
sending data to the cloud.  \cite{teerapittayanon2017distributed} shows that
the scheme reduces latency and communication cost while achieving almost the
same accuracy.  If LA is not successful, sensors send their data to the cloud
for further processing.  In this case, high communication overhead ensues.

In this paper, we consider the same scenario as in
\cite{teerapittayanon2017distributed} where there are many sensors collecting
data and transmitting them to the CN for processing (e.g. for object
classification). The LA also resides on the local area network that connects the
sensors and thus transmissions from sensors to the LA incur low communication
overhead. The LA may have sufficient computation capacity to perform the early
stages of inference tasks as in \cite{teerapittayanon2017distributed}. However,
when the LA has insufficient confidence of the inference result, sensors should send their data to the CN. In such cases, \cite{teerapittayanon2017distributed} 
suggests that all sensors send their data to th CN for further processing. In this work,
instead of
transmitting all sensor data to the CN, we propose SensorDrop, a sensor
selection approach that only transmits relevant data from a subset of sensors.
SensorDrop is motivated by the observation that distributed sensor data tends to
have high spatial correlation. Take object recognition using a multi-view
multi-camera network as an example. An object can appear in the field of view (FoV)
of multiple cameras concurrently. For the purpose of object recognition, it may
suffice to utilize the inputs from one or a subset of these cameras. Doing so
can reduce communication costs by transmitting less data to the CN.  

It is non-trivial to identify the suitable subset of sensors to transmit their
data without any expert knowledge. Such decisions are clearly input dependent
as the correlation of the data among sensors as well as their informativeness
to the target task vary with the type and location of interested events. For
instance, in the multi-view multi-camera object recognition application
discussed earlier, even among cameras that have a common object in their
FoV, the size and clarity of the object generally differ from one camera to
another.  To determine which sensors to transmit their data, in SensorDrop, we
propose an reinforcement learning-based approach to identify useful sensor data.
We adopt the Advantage Actor-Critic (A2C) reinforcement learning method, and
devise a reward function that takes into account both the accuracy of the
inference task and the communication cost. Evaluations using a multi-view
multi-camera dataset shows that SensorDrop indeed outperforms baseline methods
in communication costs with
only minor performance degradation in inference accuracy. Specifically, with over
74\% reduction in communication costs, the inference accuracy only degrades by about
10\%. We further conduct extensive experiments to investigate the impact of
parameter settings on the trade-offs between accuracy and communication costs. 

The remainder of the paper is organized as follows. In Section~\ref{sec:2}, we
review the basic concept of A2C reinforcement learning. Section~\ref{sec:3}
presents the system model and the problem definition. The proposed A2C
SensorDrop controller is discussed in Section~\ref{sec:4}. Section~\ref{sec:5}
describes the dataset and the experimental setup, and demonstrates the effectiveness of
the proposed scheme in balancing inference accuracy and communication overhead.
Finally, Section \ref{sec:6} concludes the paper.

\section{Background - Advantage Actor-Critic Based Reinforcement Learning Method (A2C)}\label{sec:2}

Reinforcement learning approaches are being used in many problems that mapping situations to actions to maximize a reward signal. 
Unlike conventional ML methods, a learner (agent) is not told which actions to take; instead, it explores different actions to maximize 
the feedback it receives from its environment 
\cite{sutton2018reinforcement}. In addition to the  agent and the environment, other major components of reinforcement learning include policy, reward and value functions. A \textit{Policy} is a rule used by the agent to decide what actions to take. \textit{Reward} is a signal that tells how good or bad agent actions are on the environment. The agent aims to maximize the cumulative reward.
\textit{Value functions} stand for the value of a state, i.e., the estimate of the expected return of being in a given state.

	\begin{figure}[!t]
		\centering
		\includegraphics[scale=0.24]{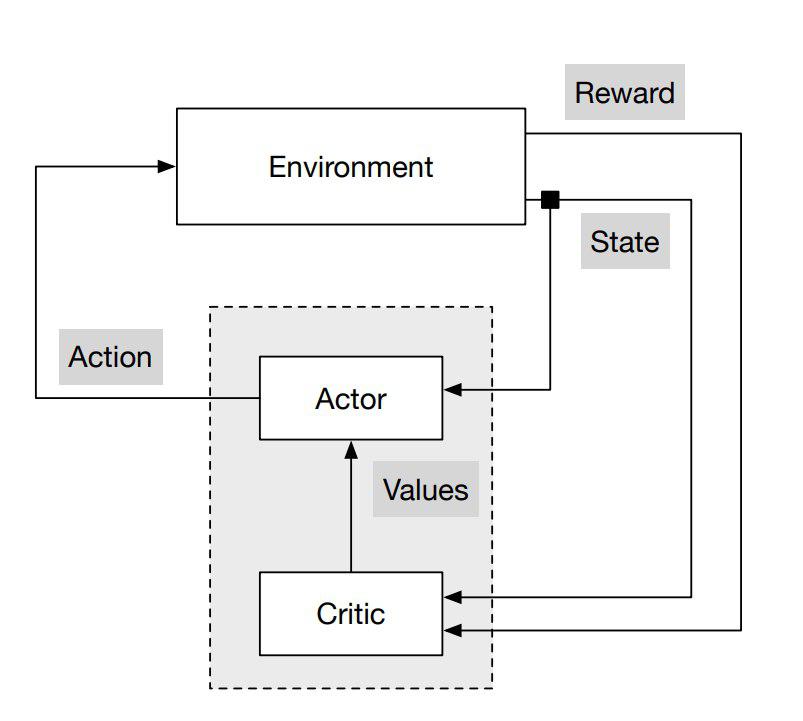}
		\caption{Actor-Critic general model}
		\label{fig:actor_critic}
	\end{figure}

Actor-critic based approaches \cite{grondman2012survey} are one of several RL algorithms exploited in different applications. A general structure of these algorithms is shown in Fig. \ref{fig:actor_critic}. 
The actor-network is responsible for deciding which actions should be taken and the critic network criticizes the actions of the actor by investigating new states and the corresponding reward value.

The actor and critic parameters are represented by $\theta$ and $W$. The actor learns the optimal policy $\pi_\theta(a|s)$ gradually while the critic learns to 
estimate the Q-function considering the value of reward $R(s,a)$. Mathematically, let $V(s')$ be the value function of next state $s'$ 
and $\gamma$ be the constant discount factor. The Q-function can be written as:
	\begin{equation}
	Q(s, a) = R (s, a)+ \gamma \cdot V(s').
	\label{eq:Q}
	\end{equation}
	$Q(s, a)$ can be further decomposed into the state-value function $V(s)$ and the advantage value $A(s, a)$. $A(s, a)$ is a measure of how much a certain action is better than other actions in a specific state, i.e.,
	$A(s,a) = Q(s, a) - V(s)$. Equivalently, we have:  
	\begin{equation}
	A(s,a) = R(s, a) + \gamma \cdot V(s') - V(s).
	\label{eq:Advantage}
	\end{equation}
	
	
	In A2C, the critic and actor weights are updated using \eqref{eq:update_critic} and \eqref{eq:update_actor} where $\beta$ and $\alpha$ are the learning rates. 
	\begin{equation}
	\nabla W = \beta A(s,a)\nabla V(s)
	\label{eq:update_critic}
	\end{equation}
	\begin{equation}
	\nabla\theta = \alpha \nabla_{\theta}(log \pi_{\theta}(a|s))  A(s,a)
	\label{eq:update_actor}
	\end{equation}
	
	
	%
	%
	%
	\begin{figure}[!t]
		\centering
		\includegraphics[scale=0.06]{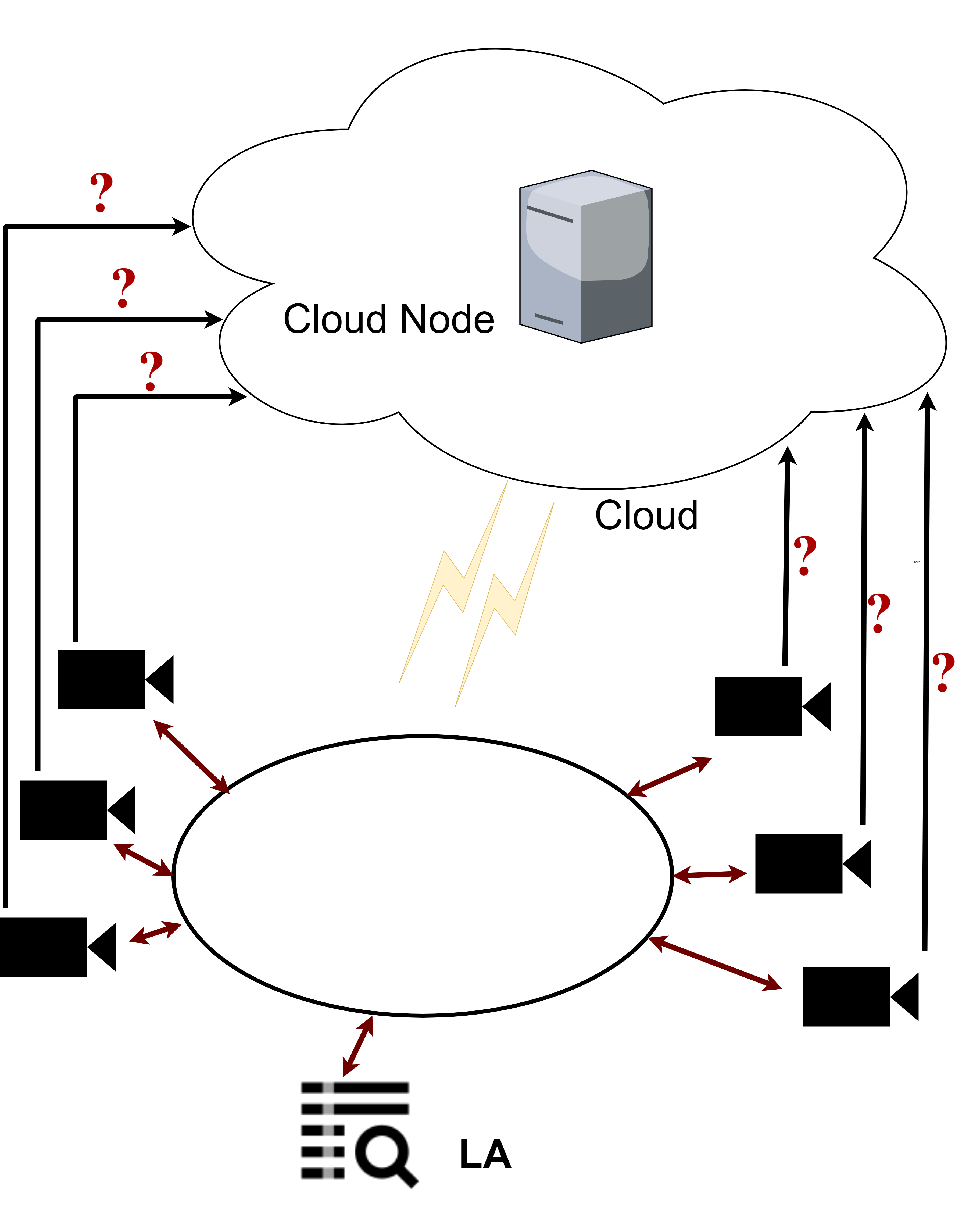}
		\caption{System model}
		\label{fig:system_model}
	\end{figure}

	\section{System Model and Problem Definition}\label{sec:3}
	
    Similar to \cite{teerapittayanon2017distributed}, we consider a system architecture that consists of three main parts: 1) end devices (Sensors) 2) an LA which is on the same network of the sensors 3) a CN in the cloud. Fig. \ref{fig:system_model} shows an overall structure of the model, 
\begin{itemize}
    \item The end devices include sensors collecting information from their environment.
    As in \cite{teerapittayanon2017distributed}, 
    we assume that sensors are low-cost and 
    capable of performing simple tasks such as limited numbers of convolutional neural network (CNN) layers. To further reduce complexity, only quantized CNNs are utilized at the sensors.
    \item     An LA is an edge device on the same network as the sensors \cite{teerapittayanon2017distributed} which can collect sensor data at low communication overhead compared to transferring data to the cloud and perform simple control tasks. LA nodes have lower computation capability than cloud nodes.
    \item     The CN is a node located in the cloud. It has high computational capabilities and  can perform complex ML tasks with significant data processing requirements. 
\end{itemize}

The network is deployed to collect data from distributed sensors and perform a complicated ML task (e.g., object recognition and tracking). It is assumed that sensors are not able  to execute the complete ML operations individually (either because they do not have sufficient processing power or the ML task requires data from many sensors).


The system architecture is well suited for IoT applications. However, one main disadvantage is the large communication cost of sending data from all sensors to the CN. To reduce the communication cost, we take advantage of correlations among data from different sensors as well as the processing power at the LA. The LA selectively instruct sensors to send data to the CN. 

LA data selection can greatly reduce the communication overhead. However, the LA has no prior knowledge of how is the degree of redundancy among sensors or how much relevant a sensor's data would be for the intended ML task at CN. Reinforcement learning methods provide a framework for learning and decision making in presence of uncertainty.


	\begin{figure*}[!t]
		\centering
		\includegraphics[scale=0.31]{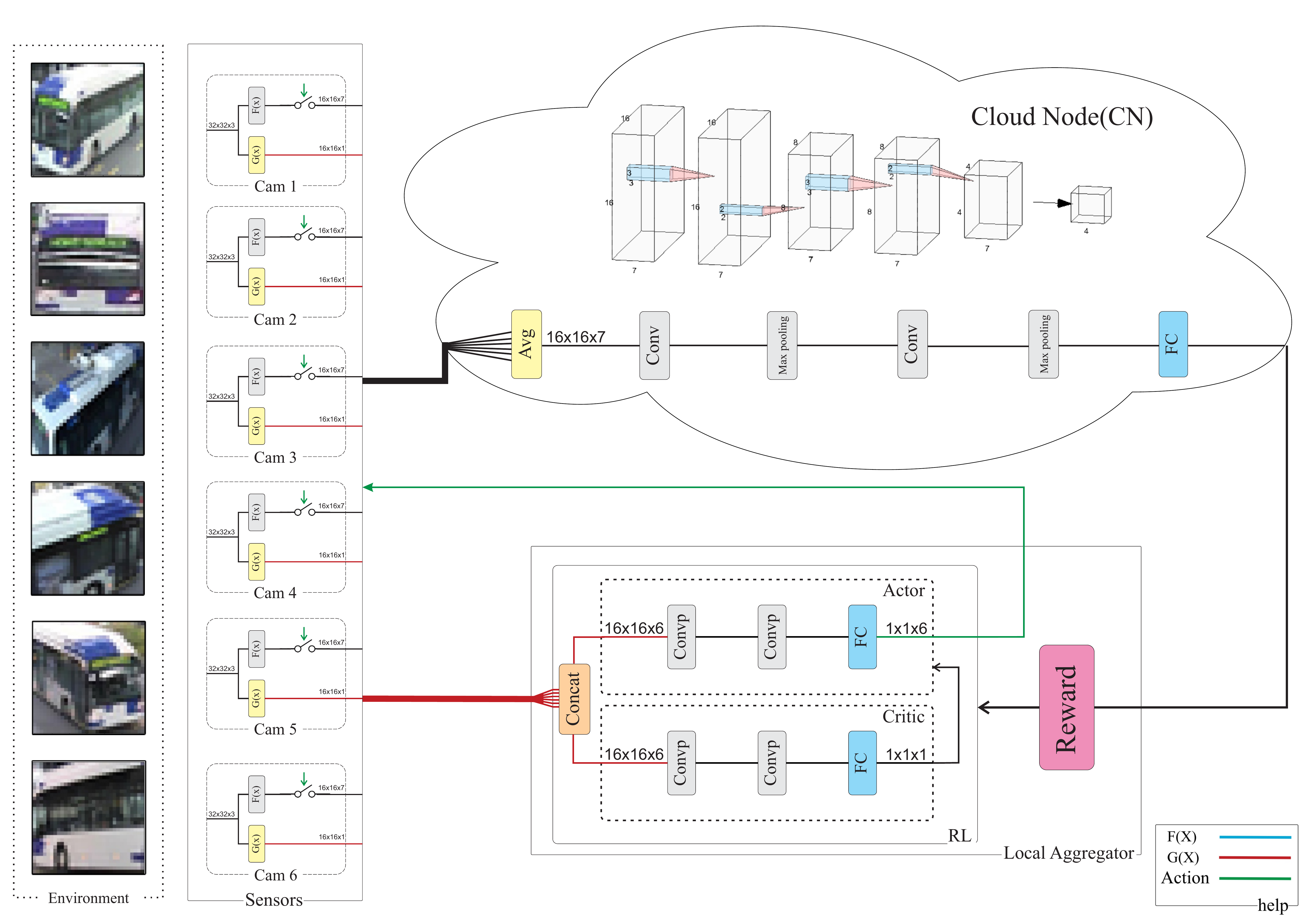}
		\caption{Detailed description of the proposed architecture}
		\label{fig:propose}	
	\end{figure*}

In this study, we utilize a neural network (NN) located at the LA to act as a controller and decide which sensor data should be transmitted.
The approach is particularly suitable in the following scenarios:
\textbf{a)} data collected by individual sensors (e.g., an image from one of the security cameras) may not contain much information for the task  performed at the cloud. For instance, the image of one camera is blurry or no object is visible in that image; 
\textbf{b)} multiple sensors report very similar data (e.g., several security cameras having similar views at the same time), and thus it suffices to send only a subset of them to the CN. 
This situation is common in IoT networks due to sparsity and locality of spatial data. 

It is worth mentioning that, compared to \cite{teerapittayanon2017distributed}, no further processing on the sensor's data is performed on the LA , rather, the LA makes binary decisions regarding whether data from a sensor should be sent to the cloud or not. In fact, our approach is orthogonal and complementary to that in \cite{teerapittayanon2017distributed}, { \color{black} i.e., it is possible to have an LA which first makes inference locally, and if it is not successful, performs the sensor-selection procedure to selectively send data of sensors to the CN. }

	
	\section{Proposed Scheme}\label{sec:4}
	
	We aim to design a controller at the LA that, given some representation of the the sensors' information, can determine the utility of forwarding a sensor's data to the CN. Consider a network of $N$ sensors. Let  $X_i$ be the raw data collected by sensor $i$ ($i\in\{1,2,\cdots,N\}$). The sensor may pass the raw information to the CN or may perform some initial processing on that (if it has enough computational power). In a general form, we use $F(X_i)$, $i\in\{1,2,\cdots,N\}$ to denote the data that the sensor transmit to the CN after such processing step (if the sensor is very simple and not able to perform any processing $F(X_i)$ could be an identity function). $F(X_i)$, if transmitted, will be utilized by the CN to perform the target ML task.

	Each sensor also has another output toward the LA using which the LA decides which sensor should send data to the CN. This output is denoted by $G(X_i)$, $i\in\{1,2,\cdots,N\}$. As the LA does not want to perform the actual ML task on the data and is responsible for sensor selection only, $G(X_i)$ is usually a low dimensional representation (e.g., a lower resolution image) of $X_i$. Noting that $G(X_i)$ could be an identity function as well but it is not a good choice as it means that we are sending all row sensors' data to the LA which is not needed in that details.

	
	The controller (located at the LA) should be trained to observe the received $ G(X_i)$ and then decide \textit{based on a metric}. As the accuracy of inference is important, the metric should retain the accuracy while reducing the overall communication-overhead. 
	Clearly, there is a tradeoff between these two factors, and based on the requirement of each application, an appropriate selection \textit{metric} should be defined.
	

	For RL modeling, we consider the combination of sensors and the CN as the \textit{environment}. In this environment, each sensor collects data $X_i$ and sends $G(X_i)$, to the LA. 
	The set of $G(X_i)'s$ received from all sensors constitutes the \textit{state} of the environment, $\mathcal{S}$.
			\begin{equation}
		{\mathcal{S} = [G(X_1) G(X_2) \cdots G(X_N)]}, 
		\label{eq:z_in}
	\end{equation}

	Based on the network \textit{state}, the RL \textit{agent}  determines the appropriate \textit{action}. In our model, the \textit{action space} is the set of $N$-dimensional binary vectors $\mathcal{V}\in \{0,1\}^N$, corresponding to which sensors should send their data to the CN and vice versa, i.e., its $i$th element determines whether the $i$th sensor's data should be dropped or transmitted. 
	$\mathcal{F}_{actor}({\mathcal{S}}, \theta)$ represents the agent neural-network equivalent function considering the environment state  $\mathcal{S}$ and $\theta$ shows the parameters of the actor network which are set during the training procedure.


	 After receiving a subset of the sensors' data, the CN performs the desirable ML task. 
	The CN first computes $\mathcal{L}$ as the average of received information,
{\color{black}
	\begin{equation}
	\label{avg_CN}
			\mathcal{L}=\sum_{i=\{1,2,\cdots,N\}}F(X_i)\cdot\frac{\mathcal{V}_i}{\left\lVert\mathcal{V}_i\right\rVert}.		
	\end{equation}}
	The averaging method at \eqref{avg_CN} can be changed based on the requirement of the application; we should just make sure that we have a method at the CN to make sure the input dimension of the CN network is constant regardless of the number of sensors selected, e.g. by averaging in this application the CN neural network does not need to know how many sensors are selected. The resulted $\mathcal{L}$ is then  passed through the CN trained neural network for classification, $\mathcal{F}_{CN}(\mathcal{L})$. Finally, the \textit{reward} of the action is determined based on the inference accuracy and communication overhead. The RL agent uses this reward to make better decisions in future steps.

	The policy we are looking for is to learn what types of data are useful for the ML task. 
	The critic network measures the quality of the controller's decision based on the classification result at the CN, updates its parameters and provides feedback to the actor about the effect of the taken action. The actor exploits this feedback to update its network weights.

To fully specify the A2C RL algorithm, we need to define the reward function that measures how good an action is.
Here we define a sample reward function that account for both the accuracy and communication overhead. 
	
	
	\begin{equation}
	Reward = \begin{cases} k_1-k_2.(\frac{d_{active}}{N})^2 & \mbox{Correct}\\-\zeta  & \mbox{Not correct}\end{cases},
	\label{eq:reward}
	\end{equation}
	where $d_{active}$ is the number of selected sensors (the number of ones in the action vector $\mathcal{V}$).
	In \eqref{eq:reward}, setting $k_1, k_2$ and $\zeta$ to different positive values, we can strike a balance between reducing communication overhead and high inference accuracy. As can be seen, if the CN misclassifies, a negative reward of $-\zeta$ will be returned to the RL agent. Otherwise, a positive reward is generated. This reward is larger when fewer sensors have transmitted their data to the CN, i.e., with less communication overhead. 
	
	Given \eqref{eq:reward}, the advantage value is calculated in \eqref{eq:Advantage} and the actor-critic weights are updated based on  \eqref{eq:update_critic} and \eqref{eq:update_actor}. 
	Alg.~\ref{alg:alg1} presents the  training procedure of the RL agent.

	Fig. \ref{fig:propose} provides the detailed network structure of the proposed scheme. \textit{We note} that the specific operations at the sensors as well as  the neural network architecture at the CN are for the particular application that we will explain in Section \ref{sec:5}. For other applications, the sensor and the CN neural nets can be modified as needed and it is independent of the proposed RL-bases sensor drop mechanism.

	\begin{algorithm}[t]
		\caption{Training RL agent for $N^{epochs}$ epochs}
		\label{alg:alg1}
		\begin{algorithmic}[1]
			\REQUIRE Input data for each of the $N$ sensors, $\mathcal{F}_{actor}({\mathcal{S}}, \theta)$, and $\mathcal{F}_{CN}(.)$ 
			\STATE Calculate $F(X_i), G(X_i)$ for all $X_i$
			\STATE Initialize parameters of actor $\theta$ and critic $w$,
			\FOR{$t=1 \leftarrow N^{epochs}$}
			\STATE ${\mathcal{S} \leftarrow [G(X_1) G(X_2) \cdots G(X_N)]}$
			\STATE $\mathcal{V} \leftarrow \mathcal{F}_{actor}({\mathcal{S}}, \theta)$,
			\STATE $\mathcal{L} \leftarrow \sum_{i=\{1,2,\cdots,N\}}F(X_i)\cdot\frac{\mathcal{V}_i}{\left\lVert\mathcal{V}_i\right\rVert}$,
			\STATE $ \mbox{\textit{final-decision}} \leftarrow \mathcal{F}_{CN}(\mathcal{L})$,
			\STATE Compare the \textit{final-decision} with the correct label and calculate reward using \eqref{eq:reward},
			\STATE Determine the advantage using \eqref{eq:Advantage},
			\STATE Update  critic and actor weights based on \eqref{eq:update_critic} and \eqref{eq:update_actor}, respectively. 
			\ENDFOR
		\end{algorithmic}
	\end{algorithm}

	%
	\section{Evaluation}\label{sec:5}
	In this section, we present the implementation and evaluation of the proposed approach. 
	
	\subsection{Dataset} 
	We use the same dataset as in \cite{teerapittayanon2017distributed} for evaluation of the proposed method. 	
	Roig \textit{et al.} \cite{roig2011conditional} first introduced this multi-view multi-camera dataset that presents multi-view of same objects. This dataset consists of images captured from 6 cameras simultaneously. 
	\cite{teerapittayanon2017distributed} further cleaned up this dataset. Similar to \cite{teerapittayanon2017distributed}, the dataset has been split to 680 training samples and 171 testing samples. It contains four classes, namely, 
	of car, bus and person and images that do not contain any of defined objects.
    
    \subsection{Implementation details} \label{ssec:imp}
    
    \textbf{Neural Network Structures:} The detailed network setup of SensorDrop for multi-view multi-camera image classification is provided on Fig. \ref{fig:propose}. Following the same terminology as in \cite{teerapittayanon2017distributed}, the basic processing block of our network is called \textit{ConvP} consisting of a convolutional layer followed by a max pooling layer.
    
    {\color{black}
    We adopt similar network topology and neural network structures at the sensors on the CN as those in \cite{teerapittayanon2017distributed}, i.e., 6 cameras (sensors) collect their images ($X_i$), and $F(X_i)$ is considered as a ConvP layer following the binary neural network (BNN) architecture \cite{courbariaux2015binaryconnect}. 
	$G(X_i)$, in this setup, is defined as the average over all channels of $F(X_i)$. 
		}

    
    The structure of the actor and critic neural networks are similar to each other, each consisting of two consecutive ConvP layers and a fully-connected layer. The CN takes the selected $F(X_i)$ and computes an average of the inputs.  
    Two sets of convolutional layer is then applied followed by a max pooling layer and a fully-connected layer with 4 outputs (a 4-dimensional one-hot vector representing four different labels available: \textit{bus, person, car} and \textit{no caption} images). The exact size of each layer 
    is given in Fig. \ref{fig:propose}.\\
    
    \textbf{Training: }
    Training of the RL-agent depends on its environment and the reward that it gets from its actions. Therefore prior to RL-agent training we need to have our environment completely setup, the neural networks at the sensors and the CN should have been trained. In our experiment, these networks are trained using typical supervised method where it is assumed that all sensors are sending their data to the CN. In this stage, Adam optimizer is used with a learning rate of 0.001 and a batch size of 50.

    Having the environment, Alg.~\ref{alg:alg1} is used for training the RL-agent where we adopt the RMSProp optimizer and the learning rate of 0.0001 for 3000 epochs. The parameters  in \eqref{eq:reward} are set as $k_1=200$, $k_2=100$, and $N=6$ as there are 6 cameras (sensors) in the multi-view multi-camera dataset.
    
	\subsection{Experimental Results}
	\begin{figure}
		\raggedleft 
		\includegraphics[scale=0.26]{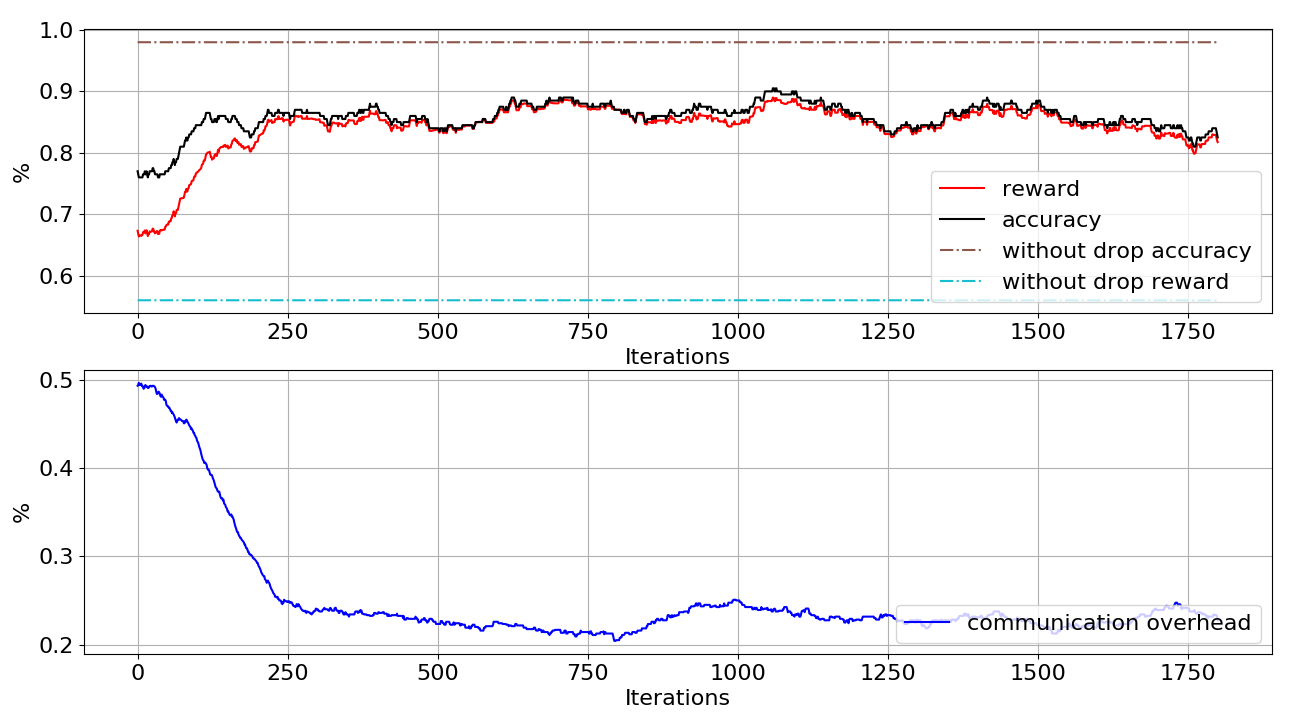}
		\caption{Convergence of the RL algorithm. Top plot: accuracy and reward vs training iterations; Bottom plot: accuracy and activity vs training iterations}
		\label{fig:resplot}
	\end{figure}

	
    We evaluate the inference accuracy of the system and the reduction in communication-overhead. \\

\subsubsection{\textbf{Convergence}}
First we investigate the convergence of Algorithm 1. In this experiments, we use  \eqref{eq:reward} (with parameters specified in Section \ref{ssec:imp}) as the reward function.   The results are presented in Fig.~\ref{fig:resplot}.  

Fig. \ref{fig:resplot} shows reward, accuracy and communication overhead over training iterations. In the plots, we normalize accuracy and communication costs using the baseline values in absence of SensorDrop. It is observed that all measures in SensorDrop are in general improve with more training iterations and we achieve higher rewards at the expense of  a small decrease in accuracy. Also as can be seen, at some points during training the accuracy  decreases as the result of incorrectly dropping sensors, but the algorithm manages to correct its decisions in the subsequent iterations.

The bottom plot of Fig. \ref{fig:resplot} reveals how effective SensorDrop is in reducing communication overhead while still maintaining accurate prediction. As can be seen, when the RL agent sees more (action, reward) tuples, it gradually learns when to drop sensors to reduce overhead without negatively  affecting the accuracy. \\


	\begin{table}[!t]
		\renewcommand{\arraystretch}{1.3}
		\caption{The Average Accuracy and Communication Overhead in different scenarios}
		\label{tab:tab1}
		\centering
		\begin{tabular}{c c c c}
			\hline
			& Accuracy &  \thead{Communication \\ Overhead} & Reward\\
			\hline
			SensorDrop - reward \eqref{eq:reward} & $\textbf{85.0}\%$ & $\textbf{25.8\%}$ & {0.84}\\
			BaseLine & ${98.6}\%$ & $100\%$ & 0.51\\
			RandomDrop & $70.3\%$ & $78.8\%$ & 0.65 \\
			\hline
			SensorDrop - reward \eqref{eq:ex_reward} & $\textbf{88.0}\%$ & $\textbf{35.2\%}$ & -- \\
			\hline
		\end{tabular}
	\end{table}
	
    \paragraph*{\textbf{Test Performance}}
	Table \ref{tab:tab1} summarizes the test performance of SensorDrop over the 171 held-out samples which have not been seen during the training phase. The first line is the result when we use the proposed method with \eqref{eq:reward} as the reward function.
	As we are not aware of any similar competing technique in literature for dropping sensor information, for comparison, we report the results of two naive methods, i.e., Baseline and RandomDrop. In the \textit{Baseline method} there is no sensor drop. 
	In \textit{RandomDrop}, to reduce communication overhead, each sensor randomly decides to send its data with probability $\rho$.  In this experiment, we set $\rho=0.75$ so that about $25\%$ of the communication  overhead can be reduced.
	{\color{black} 	Last line of Table I presents the test performance of the SensorDrop with another reward function that will be discussed in Section \ref{ssec:newreaward}.}
	
	
    In Table I, the communication overhead of all schemes are normalized by that of the Baseline.
	The baseline  has the highest communication overhead while achieving the highest accuracy as all sensor data is transmitted. 

	RandomDrop, achieves lower communication overhead at the cost of lower classification accuracy e.g., $70\%$. One important point about RandomDrop method is that this method does not have a classification accuracy, i.e., sometimes it is higher than $70\%$ and sometimes it is lower. It is due to that in some cases random dropping may results to good selection of useful data and so we get high accuracy, and sometimes it drops important images which leads to poor classification results. The communication overhead of the RandomDrop is about $78.8\%$ 
	(if we had more test samples it should be more closer to $75\%$ but now since we have limited number of test samples, it is a little bit deviated from $75\%$). 
		
	Among all, SensorDrop has the lowest communication overhead with over  $64\%$ reduction in data transmission compared with the BaseLine approach. Furthermore. since the sensor selection are based on a well-trained controller, the accuracy of the scheme is about  $85.0\%$. 

\begin{figure}[!t]
		\centering
		\includegraphics[scale=.6]{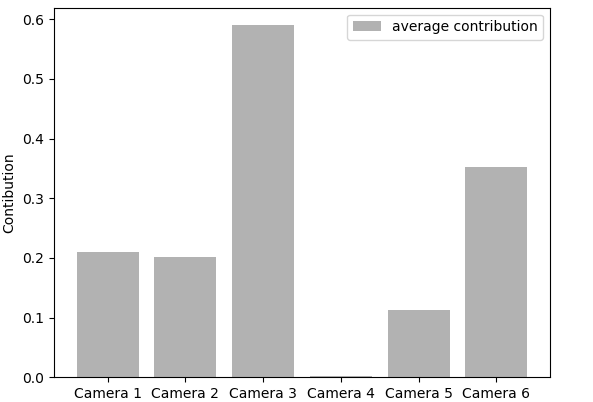}
		\caption{The average contribution of each sensor}
		\label{fig:avgcont}
	\end{figure}

	To further understand the effectiveness of SensorDrop, the average contribution of each sensor (camera) is depicted in Fig. \ref{fig:avgcont}, i.e., how often each sensor needs to transmit its data. 
	{\color{black}
	It can be seen that most of the sensors send a small percentage of the total data they collected with the except of camera 3, which contributes most data.}
	
	It is worth mentioning that the outputs of SensorDrop provide insights on the deployment of the sensors. For instance, in the security camera dataset, data from camera 3 and 6 are important for correct inference, while in contrast camera 4 and 5 do not contribute much. This implies that camera 4 and 5 are positioned in suboptimal locations. We might decide to relocate them or shut them down to conserve resources. 

	\begin{figure}[!t]
		\centering
		\includegraphics[scale=.55]{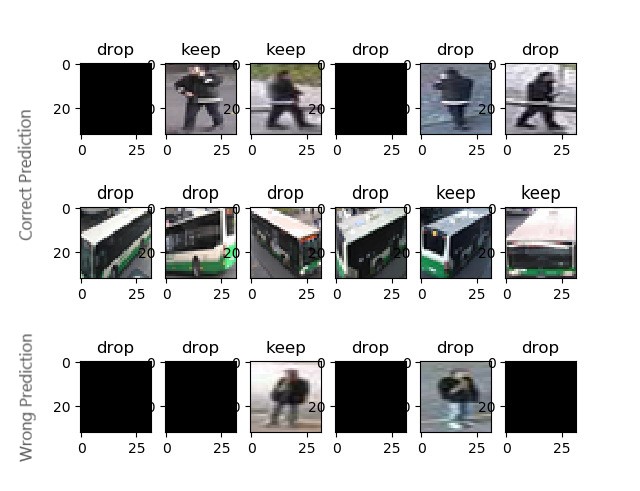}
		\caption{Qualitative evaluation of SensorDrop}
		\label{fig:qualitive}
	\end{figure}	
	
	To further gain insights on the behavior of the algorithm, 
	Fig. \ref{fig:qualitive} shows three camera views, each containing six images from the six cameras. Two of the sequences result in correct classification and one gives incorrect classification. Also included in the figure are the decisions by the RL agent in SensorDrop.
	

	In Fig. \ref{fig:qualitive}, the first row contains images from different cameras when a person is in the view of 4 cameras but not in the view of the remaining 2 cameras (as indicated by the blank ones). As expected, both of the blank images are dropped correctly, and two of the non-black images were selected to be transmitted to the CN by SensorDrop. The CN  correctly classifies the target with the two images.
	
	In the second row of Fig. \ref{fig:qualitive}, there are 6 images of a bus from different views. Clearly, there exist redundancy among these images and thus it is expected that not all images are needed for the inference task (i.e., to identify the presence of a bus). 
	 It is observed that the RL method act reasonably by selecting the last two images which results in correct classification.

	{\color{black}
	The last row of Fig. \ref{fig:qualitive} gives an example that SensorDrop failed. In this example a pedestrian is in the view of only two out of six cameras. In this case, SensorDrop selected the input from a single camera, which results in a wrong classification at the CN. Note that such events are infrequent as the overall accuracy of the model is very high as reported in Table 1.}


	\begin{figure}
		\centering 
		\includegraphics[scale=0.24]{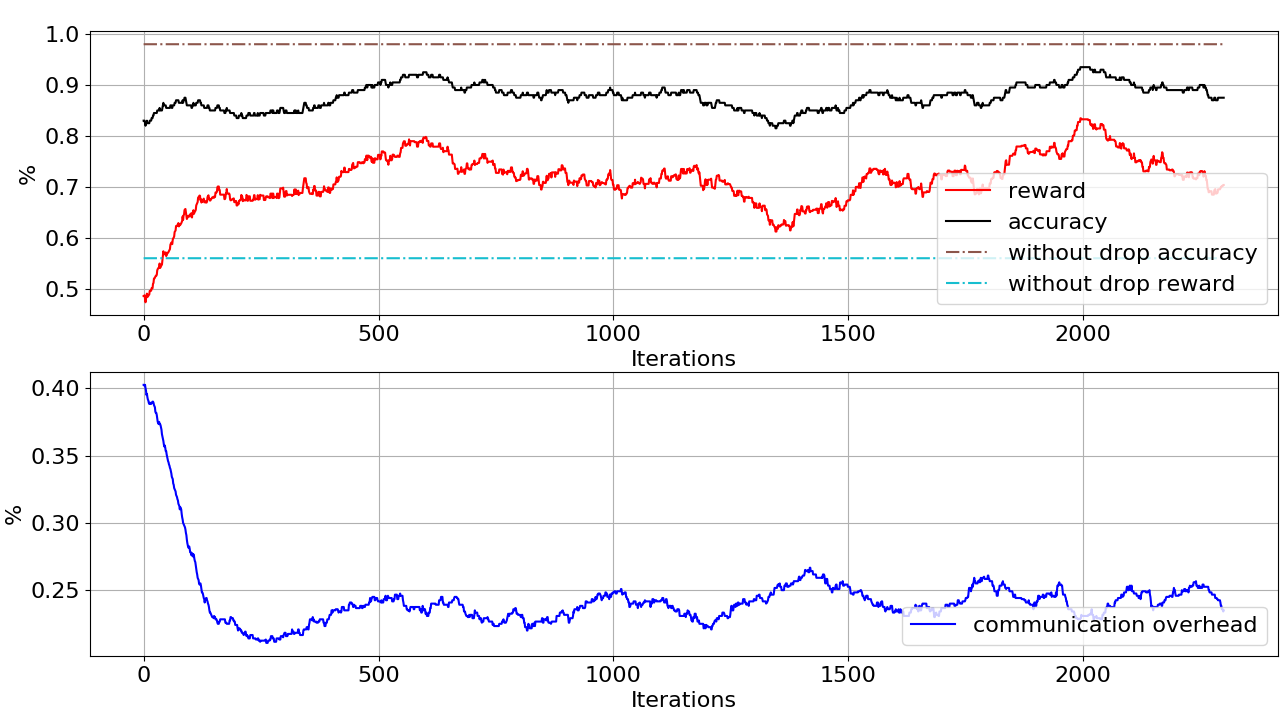}
		\caption{Convergence of the RL algorithm with Reward function of \eqref{eq:ex_reward}. Top plot: accuracy and reward vs training iterations; Bottom plot: accuracy and activity vs training iterations}
		\label{fig:resplot_reward2}
	\end{figure}

 \begin{figure}[!t]
 	\centering
 	\includegraphics[scale=0.25]{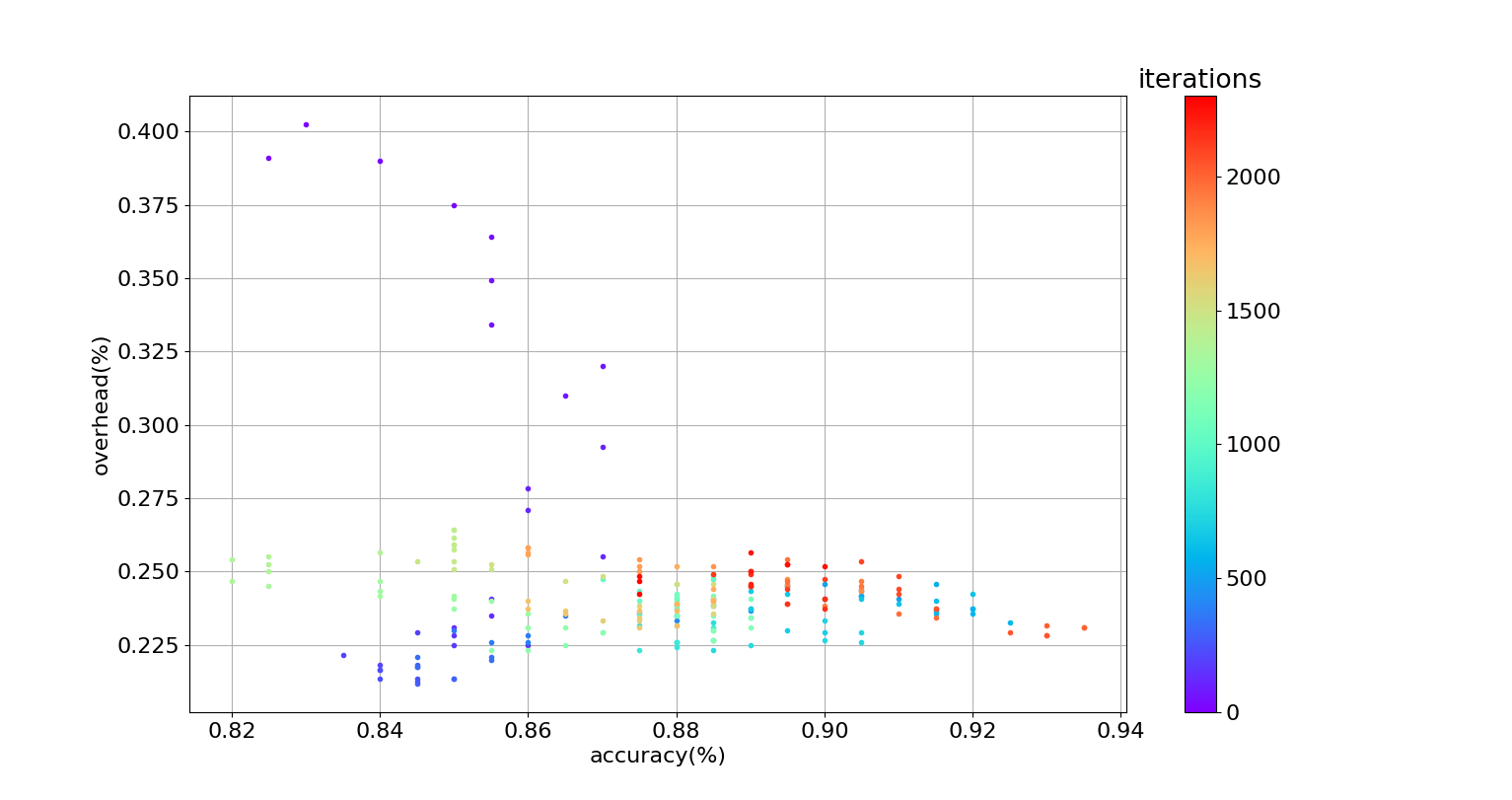}
 	\caption{Convergence of the proposed scheme with \eqref{eq:ex_reward} as the reward function}
 	\label{fig:ext_r_conv}
 \end{figure}
		
 	\subsubsection{\textbf{Trade-off between Accuracy and communication overhead}}
 	\label{ssec:newreaward}
 
	 As the final set of experiments, we explore the tradeoff between accuracy and communication overhead by adjusting the parameters in \eqref{eq:reward}.
 
	 As discussed in Section \ref{sec:4}, {\color{black} the suitable reward function is problem dependent and should be defined based on the requirements of the specific problem. We have already demonstrated the effectiveness of SensorDrop when \eqref{eq:reward} is used as the reward function. To show SensorDrop works with other rewards, we experiment with another simple reward function as,}
 \begin{equation}
 Reward = \begin{cases} K + \frac{1-K}{d_{active}} & \mbox{Correct}\\-\zeta^{'}  & \mbox{Not correct}\end{cases},
 \label{eq:ex_reward}
 \end{equation}	
 where $d_{active}$ is the number of active sensors (the number of ones in the action vector $\mathcal{V}$) and parameter $K$ is a real number in the range of $[0, 1]$, which can be tuned to give more importance to accuracy or communication overhead. A smaller $K$ will lead to greater reduction on the communication overhead, and vice versa. The other parameter $-\zeta^{'}$ is a negative reward value that the agent receives if a wrong prediction occurs in the experiments, $\zeta^{'}$ is set to  $0.75$.

 \begin{figure}[!t]
 	\centering
 	\includegraphics[scale=.18]{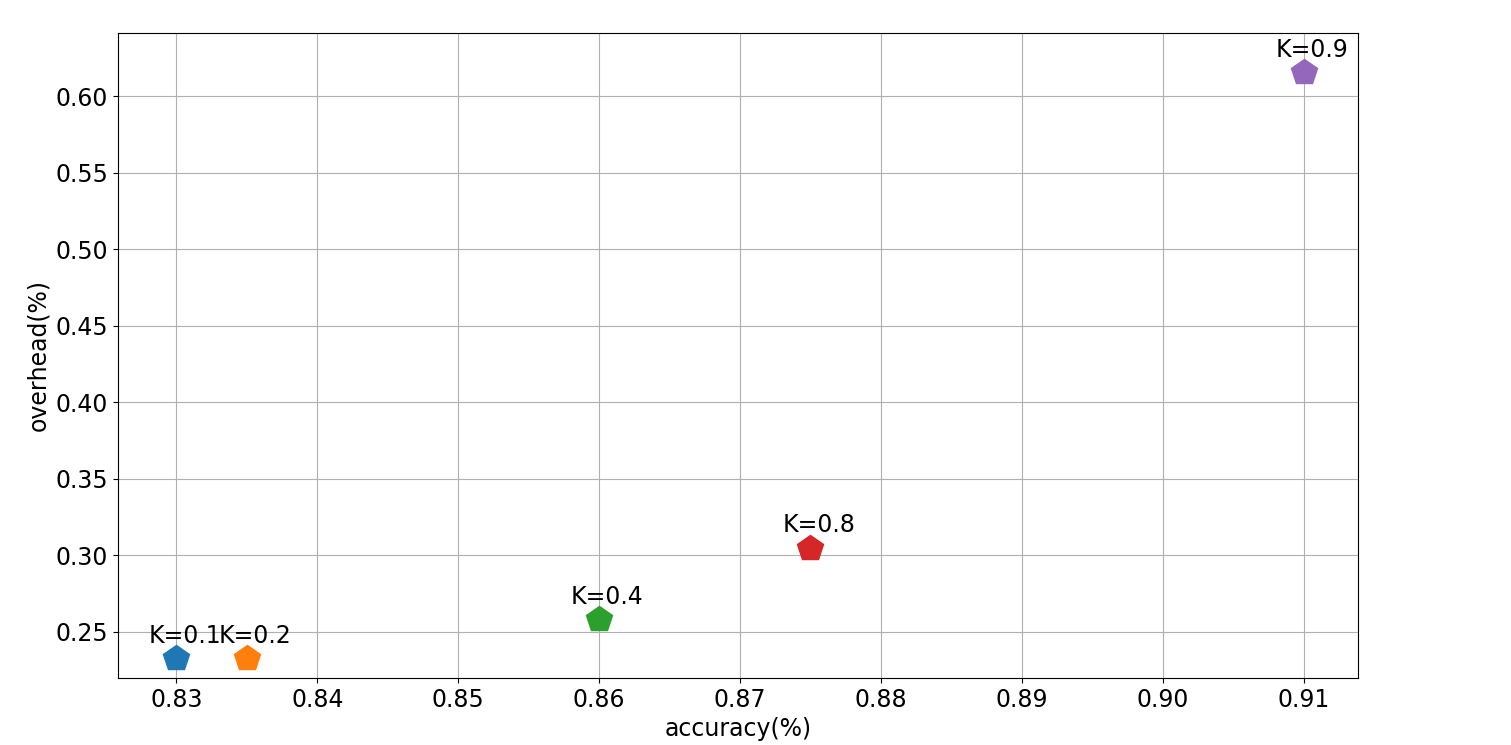}
 	\caption{ Accuracy - Communication overhead Tradeoff }
 	\label{fig:dots}
 \end{figure}  

 We trained our SensorDrop with the new reward function and $K=0.4$.
Figure \ref{fig:resplot_reward2} shows the convergence of SensorDrop.
Specially, it plots training accuracy and normalized communication overhead (over a baseline without dropping). It can be observed that the model learns to remove redundancy in the data and reduce the communication overhead while maintaining the highly accurate object detection. The \textit{test performance}  of this setting is reported as the last line of Table~I. As can be seen, the new reward function has a bit higher accuracy compared to the initial reward function in \eqref{eq:reward}, but the saving on the communication overhead is about $10\%$ lower. We did not include the reward value for this case in the table, as here we use different reward functions compared  with the other methods of Table~I.

Figure \ref{fig:resplot_reward2} results are differently illustrated in Fig.~\ref{fig:ext_r_conv}. In this figure, there are several dots showing the performance (accuracy and overhead) of the proposed method when it is trained over time. The color of each dot represent show long the network was under training (color goes from blue to red), i.e., blue dots represent the network performance in early training stages and red dots are associated to the results after many iterations. The axis of Fig. \ref{fig:ext_r_conv} are the accuracy and communication overhead of the SensorDrop scheme at that particular iteration. As can be seen, the model learns to go from high-overhead to low-overhead while maintain the high accuracy of object detection.

  Lastly, we study how parameter $K$ in \eqref{eq:ex_reward} affects the trade-off between accuracy and communication costs. In this set of experiments,$K$ varies from 0.1 to 0.9.
  The test accuracy and communication overhead are depicted in Fig. \ref{fig:dots}. 
 
 When $K=0.9$, the model learns to attain a high accuracy of $91\%$ but it incurs a higher  communication overhead. On the other end, smaller values of $K$, such as $K=0.1$, lead to significant reduction of communication overhead (to about 25\%) but at the expense of lower accuracy. It should be noted that despite the drop in accuracy for smaller $K$, the reduction is moderate since SensorDrop intelligently selects which sensors' data to be dropped.

 \section{Conclusion}\label{sec:6}    
 In this paper, We investigate the problem of reducing the communication overhead from distributed sensors to a cloud node for complex inference tasks. Designed a controller selects a subset of sensors to send data to the cloud, while keeping the accuracy at an acceptable level. Considering the dynamics of the data collected at the sensors, we devised a Advantage Actor-Critic based RL scheme to train the controller. The performance of SensorDrop has been evaluated in different settings using a real-world multi-view camera dataset. SensorDrop was shown to greatly outperform naive schemes such as no-drop and random drop. We have also demonstrated how the parameters of the RL reward function can be tuned to make an appropriate tradeoff between the accuracy and communication overhead.


	
	%

	

	%
	%

	\ifCLASSOPTIONcaptionsoff
	\newpage
	\fi

	
	
	\bibliographystyle{IEEEtran}
	\bibliography{References}	

\begin{thebibliography}{10}
\providecommand{\url}[1]{#1}
\csname url@samestyle\endcsname
\providecommand{\newblock}{\relax}
\providecommand{\bibinfo}[2]{#2}
\providecommand{\BIBentrySTDinterwordspacing}{\spaceskip=0pt\relax}
\providecommand{\BIBentryALTinterwordstretchfactor}{4}
\providecommand{\BIBentryALTinterwordspacing}{\spaceskip=\fontdimen2\font plus
\BIBentryALTinterwordstretchfactor\fontdimen3\font minus
  \fontdimen4\font\relax}
\providecommand{\BIBforeignlanguage}[2]{{%
\expandafter\ifx\csname l@#1\endcsname\relax
\typeout{** WARNING: IEEEtran.bst: No hyphenation pattern has been}%
\typeout{** loaded for the language `#1'. Using the pattern for}%
\typeout{** the default language instead.}%
\else
\language=\csname l@#1\endcsname
\fi
#2}}
\providecommand{\BIBdecl}{\relax}
\BIBdecl

\bibitem{gambirovza2018big}
J.~{\v{C}}. Gambiro{\v{z}}a and T.~Masteli, ``Big data challenges and
  trade-offs in energy efficient internet of things systems,'' in \emph{26th
  International Conference on Software, Telecommunications and Computer
  Networks (SoftCOM)}.\hskip 1em plus 0.5em minus 0.4em\relax IEEE, 2018, pp.
  1--6.

\bibitem{wu2016quantized}
J.~Wu, C.~Leng, Y.~Wang, Q.~Hu, and J.~Cheng, ``Quantized convolutional neural
  networks for mobile devices,'' in \emph{Proceedings of the IEEE Conference on
  Computer Vision and Pattern Recognition}, 2016, pp. 4820--4828.

\bibitem{huang2018data}
Z.~Huang and N.~Wang, ``Data-driven sparse structure selection for deep neural
  networks,'' in \emph{Proceedings of the European Conference on Computer
  Vision (ECCV)}, 2018, pp. 304--320.

\bibitem{wu2018blockdrop}
Z.~Wu, T.~Nagarajan, A.~Kumar, S.~Rennie, L.~S. Davis, K.~Grauman, and
  R.~Feris, ``Blockdrop: Dynamic inference paths in residual networks,'' in
  \emph{Proceedings of the IEEE Conference on Computer Vision and Pattern
  Recognition}, 2018, pp. 8817--8826.

\bibitem{teerapittayanon2016branchynet}
S.~Teerapittayanon, B.~McDanel, and H.~Kung, ``Branchynet: Fast inference via
  early exiting from deep neural networks,'' in \emph{2016 23rd International
  Conference on Pattern Recognition (ICPR)}.\hskip 1em plus 0.5em minus
  0.4em\relax IEEE, 2016, pp. 2464--2469.

\bibitem{teerapittayanon2017distributed}
S.~{Teerapittayanon}, B.~{McDanel}, and H.~T. {Kung}, ``Distributed deep neural
  networks over the cloud, the edge and end devices,'' in \emph{2017 IEEE 37th
  International Conference on Distributed Computing Systems (ICDCS)}, June
  2017, pp. 328--339.

\bibitem{sutton2018reinforcement}
R.~S. Sutton and A.~G. Barto, \emph{Reinforcement learning: An
  introduction}.\hskip 1em plus 0.5em minus 0.4em\relax MIT press, 2018.

\bibitem{grondman2012survey}
I.~Grondman, L.~Busoniu, G.~A. Lopes, and R.~Babuska, ``A survey of
  actor-critic reinforcement learning: Standard and natural policy gradients,''
  \emph{IEEE Transactions on Systems, Man, and Cybernetics, Part C
  (Applications and Reviews)}, vol.~42, no.~6, pp. 1291--1307, 2012.

\bibitem{roig2011conditional}
G.~Roig, X.~Boix, H.~B. Shitrit, and P.~Fua, ``Conditional random fields for
  multi-camera object detection,'' in \emph{2011 International Conference on
  Computer Vision}.\hskip 1em plus 0.5em minus 0.4em\relax IEEE, 2011, pp.
  563--570.

\bibitem{courbariaux2015binaryconnect}
M.~Courbariaux, Y.~Bengio, and J.-P. David, ``Binaryconnect: Training deep
  neural networks with binary weights during propagations,'' in \emph{Advances
  in neural information processing systems}, 2015, pp. 3123--3131.

\end{thebibliography}
\end{document}